\documentclass[twocolumn]{aastex631}
\usepackage{color}
\usepackage[titletoc]{appendix}
\usepackage{amsmath}
\usepackage{amssymb}
\usepackage{mathtools}
\usepackage{upgreek}
\usepackage{float}
\usepackage{comment}
\usepackage{enumitem}
\usepackage{natbib}
\usepackage{graphicx}
\usepackage{bm}
\usepackage{totcount}
\usepackage{multirow}
\usepackage{hyperref}
\usepackage{cleveref}
\usepackage{tabularx}
\usepackage[T1]{fontenc}

\newtotcounter{citnum} %From the package documentation
\def\oldbibitem{} \let\oldbibitem=\bibitem
\def\bibitem{\stepcounter{citnum}\oldbibitem}
\defcitealias{2022AJ....163..201G}{GB22}

\shortauthors{Liveoak \& Millholland}
\shorttitle{Understanding LHS 3154 b}

\begin{document} 

\title{Formation of Close-in Neptunes Around Low-Mass Stars Through Breaking Resonant Chains}

\author{Donald Liveoak}
\affiliation{Department of Physics, Massachusetts Institute of Technology, Cambridge, MA 02139, USA}
\affiliation{MIT Kavli Institute for Astrophysics and Space Research, Massachusetts Institute of Technology, Cambridge, MA 02139, USA}

\author[0000-0003-3130-2282]{Sarah C. Millholland}
\affiliation{Department of Physics, Massachusetts Institute of Technology, Cambridge, MA 02139, USA}
\affiliation{MIT Kavli Institute for Astrophysics and Space Research, Massachusetts Institute of Technology, Cambridge, MA 02139, USA}
\email{sarah.millholland@mit.edu}

\begin{abstract}
Conventional planet formation theories predict a paucity of massive planets around small stars, especially very low-mass ($0.1 - 0.3 \ M_{\odot}$) mid-to-late M dwarfs. Such tiny stars are expected to form planets of terrestrial sizes, but not much bigger. However, this expectation is challenged by the recent discovery of LHS 3154 b, a planet with period of 3.7 days and minimum mass of $13.2 \ M_{\oplus}$ orbiting a $0.11 \ M_{\odot}$ star. Here, we propose that close-in Neptune-mass planets like LHS 3154 b formed through an anomalous series of mergers from a primordial compact system of super-Earths. We perform simulations within the context of the ``breaking the chains'' scenario, in which super-Earths initially form in tightly-spaced chains of mean-motion resonances before experiencing dynamical instabilities and collisions. Planets as massive and close-in as LHS 3154 b ($M_p \sim 12 - 20 \ M_{\oplus}$, $P < 7$ days) are produced in $\sim$1\% of simulated systems, in broad agreement with their low observed occurrence. These results suggest that such planets do not require particularly unusual formation conditions but rather are an occasional byproduct of a process that is already theorized to explain compact multi-planet systems. Interestingly, our simulated systems with LHS 3154 b-like planets also contain smaller planets at around $\sim 30$ days, offering a possible test of this hypothesis.
\end{abstract}

\section{Introduction}
\label{sec: Introduction}

The formation of massive planets around very low-mass stars presents a challenge to contemporary planet formation theories. Compared to Solar-type stars, M dwarfs have longer dynamical timescales and smaller disk surface densities, such that predictions from core accretion theory have long suggested that giant planets should be very rare around M stars \citep[e.g.][]{2004ApJ...612L..73L, 2005ApJ...626.1045I}. While these predictions have been shown to be accurate to a large extent, the observational data paints a somewhat complicated picture that motivates deeper study. 

Occurrence rate studies have shown that giant planets are indeed rarer around M dwarfs than FGK stars \citep{2006ApJ...649..436E, 2007ApJ...670..833J, 2013A&A...549A.109B, 2021A&A...653A.114S, 2023AJ....165...17G, 2023MNRAS.521.3663B}. They are not entirely absent though; there are now about $\sim 10$ nontransiting and $\sim 15$ transiting short-period gas giant planets around M dwarfs \citep[e.g.][]{2022MNRAS.511...83G, 2023AJ....166...30C, 2023AJ....165..120K, 2024A&A...683A.166A, 2024AJ....167..161K}. However, most of these discovered planets are around early-type M stars. Giant planets are even rarer around mid-to-late M dwarfs. Recently, \cite{2023AJ....166...11P} conducted a volume-complete survey of 200 low-mass ($0.1-0.3 \ M_{\odot}$) M dwarfs, finding no detections and deriving a 95\% confidence upper limit of 1.5\% on the occurrence of $M_p \sin i > 1 \ M_{\mathrm{Jup}}$ giant planets out to the water snow line. 

Focusing on these very low-mass ($\lesssim 0.3 \ M_{\odot}$) M stars, it is not just Jovian planets that are rare. Such low-mass stars are expected to form planets with masses up to super-Earth scales ($\sim 5 \ M_{\oplus}$) but not much more massive \citep{2020MNRAS.491.1998M}. Indeed, it has been found that terrestrial planets are common around low-mass M dwarfs while larger planets are rare \citep[e.g.][]{ 2021A&A...653A.114S, 2023AJ....165..265M}. It thus came as a surprise when \cite{2023Sci...382.1031S} presented the radial velocity discovery of LHS 3154 b, a planet with minimum mass $13.2 \ M_{\oplus}$ in a 3.7 day orbit around a $0.11 \ M_{\odot}$ star. The planet-to-star mass ratio, $3.5\times10^{-4}$, is highest of any planet with $P < 100$ days orbiting a star with $M_{\star} < 0.25 \ M_{\odot}$ \citep{2023Sci...382.1031S}. Although it is not a Jovian, LHS 3154 b is still difficult to explain with planet formation theories. For instance, \cite{2023Sci...382.1031S} used core accretion simulations to show that a Neptune-mass planet like LHS 3154 b is only produced if the protoplanetary disk dust mass is about ten times greater than typically assumed for a $\sim0.1 \ M_{\odot}$ star.

One way that this problem might be circumvented is if the planet had an anomalous merger history. Here we explore the hypothesis that LHS 3154 b is a product of a set of collisions and mergers of a primordial compact system of Earth/super-Earth-sized planets. This scenario is motivated by the existence of compact systems of small rocky planets around very low-mass M dwarfs, such as the prototypical TRAPPIST-1 system containing seven roughly Earth-mass planets \citep{2017Natur.542..456G}. The TRAPPIST-1 planets form a chain of mean-motion resonances (MMRs) \citep[e.g.][]{luger2017seven, grimm2018nature, 2021PSJ.....2....1A}. Although resonant chains are quite rare in the observed planet sample, it is commonly believed that they were prevalent at the early stages of planet formation due to migration from planet-disk interactions \citep[e.g.][]{2006A&A...450..833C, 2007ApJ...654.1110T, 2009ApJ...699..824O, 2016MNRAS.457.2480C}. The so-called ``breaking the chains'' model posits that most of these systems underwent dynamical instabilities and collisions after disk dispersal, resulting in orbital architectures that closely resemble the properties of the observed compact multi-planet systems \citep{2017MNRAS.470.1750I, 2019MNRAS.486.3874C, 2021A&A...650A.152I, 2022MNRAS.509.2856E, 2022AJ....163..201G, 2023ASPC..534..863W, 2024arXiv240810206L}. This model is also consistent with recent findings that resonant planets are younger \citep{2024arXiv240606885D} and puffier \citep{2024A&A...687L...1L} on average. 

As a natural consequence of the ``breaking the chains'' scenario, we may expect an occasional extreme set of collisions that produces a particularly massive remnant planet. Rather than a final system of multiple close-in super-Earths/sub-Neptunes, the outcome could be a Neptune-mass planet or even larger. In this paper, we simulate the ``breaking the chains'' scenario for initially small planets around very low-mass stars and examine whether planets like LHS 3154 b are sometimes formed as a byproduct. We model our approach on \cite{2022AJ....163..201G}, hereafter \citetalias{2022AJ....163..201G}, who showed that instabilities of initially resonant systems are consistent with the intra-system mass and spacing uniformity of compact multi-planet systems.      

The paper is organized as follows. In Section \ref{sec: methods}, we describe how we create pre-instability systems of small planets, migrate them into resonant chains, and trigger instabilities and mergers. In Section \ref{sec: high-mass sample}, we first validate our simulations by comparing to \citetalias{2022AJ....163..201G} for the case of a solar-mass host star. In Section \ref{sec: low-mass sample}, we turn to the regime of small planets orbiting very low-mass stars and examine the post-instability statistics. We examine the remnant planets with the highest masses and compare their properties to LHS 3154 b. In Section \ref{sec: discussion}, we discuss our model assumptions, long-term tidal evolution, and other possible formation pathways. We conclude in Section \ref{sec: conclusion}. 

\section{Methods}
\label{sec: methods}

We model the formation of close-in planets by migrating them into resonant chains, which then undergo instabilities due to disk dissipation and atmospheric mass loss. We consider systems starting with $N = 11$ planets, which is roughly twice the inferred mean intrinsic multiplicity of observed planetary systems \citep{10.1093/mnras/sty3463, 2019MNRAS.490.4575H} and consistent with \citetalias{2022AJ....163..201G}'s approach. The planet masses are drawn from a Gaussian distribution with average mass $\overline{m}$ and standard deviation $\sigma_m$ (which we will vary). The average intra-system mass dispersion of the full population is quantified using the metric introduced in \citetalias{2022AJ....163..201G}: 
\begin{equation}
    \mathcal{D} = \frac{1}{N_{\text{sys}}} \sum_{i=1}^{N_{\text{sys}}} \frac{\sigma_{m,i}}{\overline{m}_i},
\end{equation}
where $\overline{m}_i$ and $\sigma_{m,i}$ are the mean and standard deviation of the planet masses in the $i^\text{th}$ system. 

\subsection{Resonant chain formation}
In order to generate compact systems that are likely to undergo instabilities, we create resonant chains with adjacent planet pairs in 4:3 MMRs \citepalias{2022AJ....163..201G}. We note that this set-up is an oversimplification in the sense that we generally expect a variety of first-order and second-order resonances. \cite{burn2021new} used the Generation III Bern model of planetary population synthesis \citep{2021A&A...656A..69E,
2021A&A...656A..70E} to study planet formation around low-mass stars. For a $0.1 \ M_{\odot}$ star, the most common resonances that formed were the 3:2 and 4:3, followed by the 5:4 and 2:1. The 4:3 resonance falls near the middle of this set of common MMRs, and we also require tight resonances in order for the instabilities to develop, thus justifying our choice to use the 4:3 MMR. The resulting collision outcomes will not be substantially affected by this set-up. We initialize planets in circular and coplanar orbits with period ratios of 1.37, which is slightly wide of the intendend 4:3 resonance. We do this by specifying the initial semi-major axis of the innermost planet and choosing each subsequent semi-major axis such that it conforms with the intended 1.37 period ratio. Mean anomalies are initialized randomly within $[0^{\circ}, 360^{\circ}]$.

The planets first undergo a 100 kyr period of disk-driven migration towards their host star using the \texttt{WHfast} integrator in \texttt{rebound} \citep{rebound, reboundwhfast, wisdom1991symplectic}. The integration timestep is chosen to be less than $1/15$ of the innermost planet's orbital period. The migration is modeled by exponential damping of each planet's semi-major axis and eccentricity, with characteristic timescales $\tau_a$ and $\tau_e$, given by
\begin{align}
    \tau_a &= -\frac{2\times 10^5 \text{ yr}}{\log_{10}(a_{j,0} / a_s)}& \tau_e &= -\frac{2\times 10^2 \text{ yr}}{(a_{j,0} / \text{AU})},
\end{align}
where $a_{j,0}$ is the initial semi-major axis of the $j^\text{th}$ planet and $a_s$ is a scale distance. Note that $a_{1,0}$ and $a_s$ are input parameters of our model. The exponential damping is implemented using \texttt{reboundx} \citep{reboundx}. 

\begin{figure}
    \centering
    \includegraphics[scale=0.5]{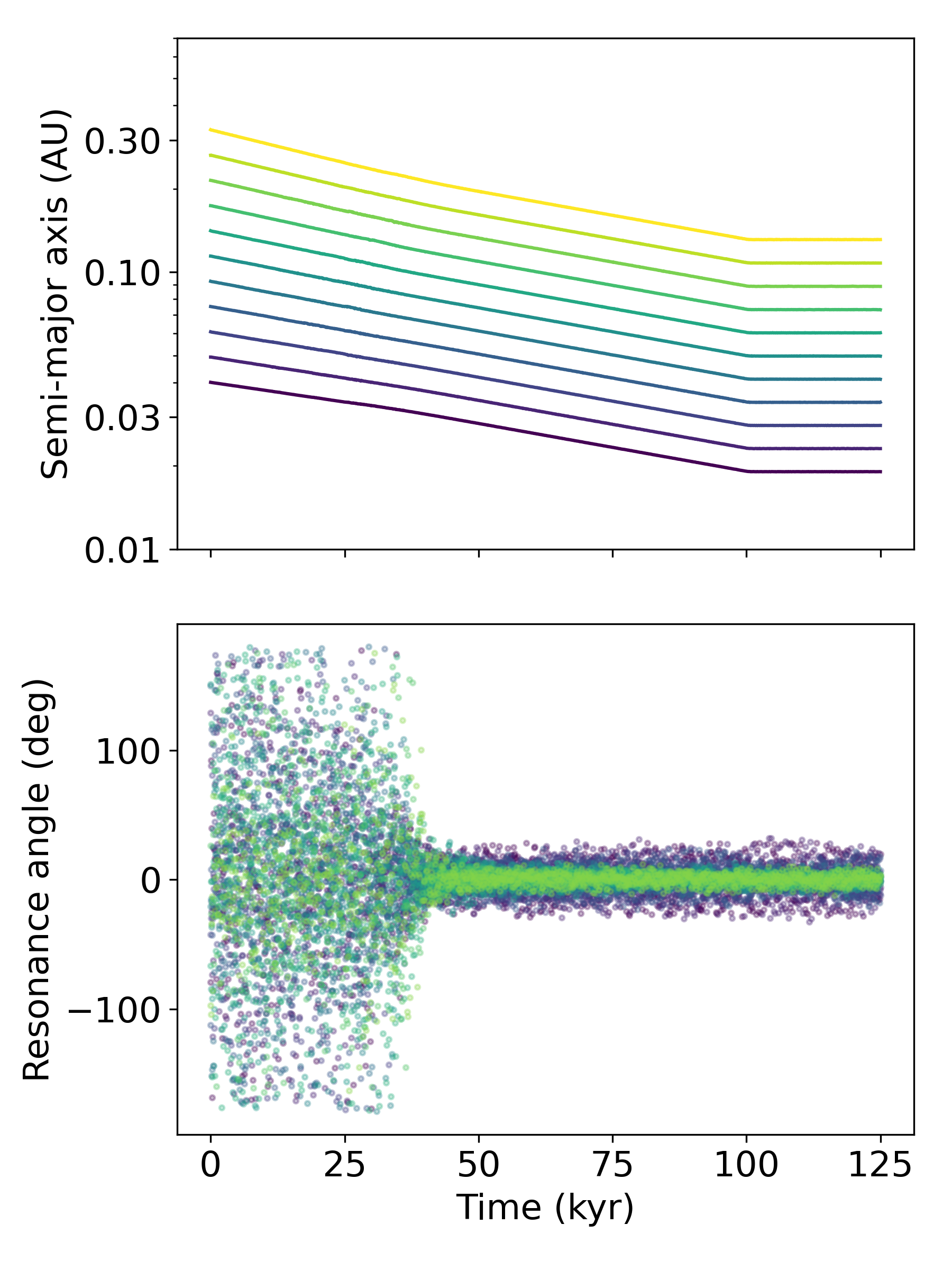}
    \caption{Top: Semi-major axes of planets in a typical system during the migration period. Bottom: Critical resonance angles of a representative system where the planets were all captured into 4:3 MMRs. Each color represents a pair of planets in the system. The innermost planet is initialized at 0.04 AU and migrates to $\sim$0.02 AU.}
    \label{fig:resonant_angles}
\end{figure}

After the migration ends, the disk is dissipated via an exponential damping of $\tau_a$ and $\tau_e$ over 10 kyr. As noted by \citetalias{2022AJ....163..201G}, this damping timescale does not significantly alter the properties of the resulting system. The system is then simulated for another 15 kyr to confirm resonance capture. Systems that undergo instabilities during disk migration or dissipation ($\sim5\%$) are discarded and not considered in the subsequent analysis. The top panel of \Cref{fig:resonant_angles} shows the semi-major axis evolution of a representative case during this first phase of the evolution.  

\begin{figure}
    \centering\includegraphics[width=0.5\textwidth]{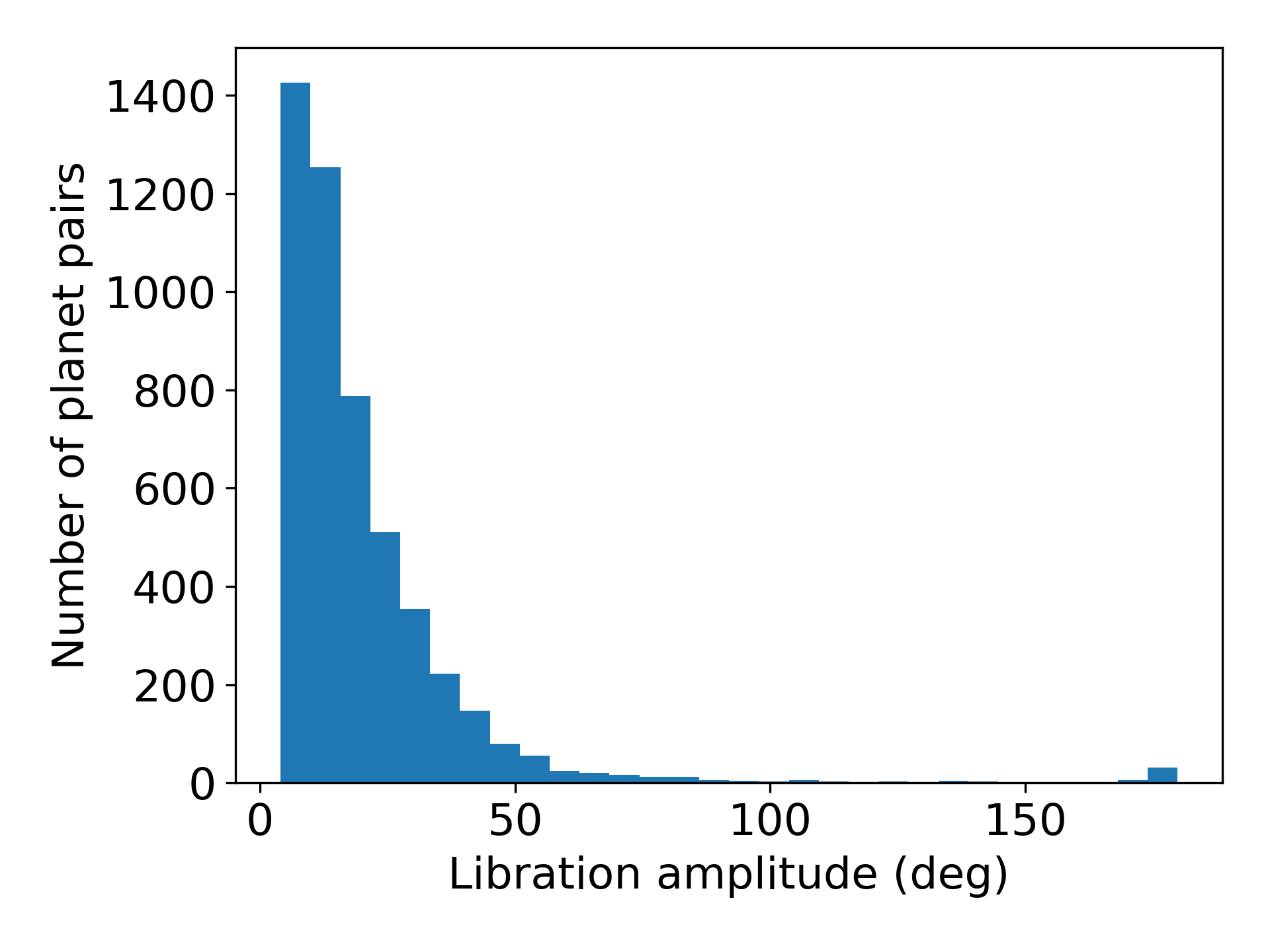}
    \caption{Distribution of libration amplitudes of the critical resonance angles in the period during and after disk dispersal.}
    \label{fig:resonance_distribution}
\end{figure}

Resonance capture is confirmed by monitoring the critical resonance angle $\theta_i = 4\lambda_{i+1}-3\lambda_i - \varpi_i$ for the 15 kyr period immediately following disk dispersal (bottom panel of \Cref{fig:resonant_angles}). Each pair of planets is considered to be captured into the intended 4:3 resonance if their period ratio is within $0.02$ of 4:3 and the post-disk dispersal libration amplitude is no more than $50^{\circ}$. Using this criteria, we report that for all samples to be defined later, over 95\% of planet pairs are captured into the intended resonance. The remaining proportion either failed to be in resonance or were captured into a nearby resonance (e.g. 7:5.)

\subsection{Collisional growth}
With stable resonant chains formed, we next excite dynamical instabilities with the goal of generating collisions and mergers. This so-called ``breaking the chains'' model has been shown to successfully reproduce many aspects of the short-period planet population \cite[e.g.][]{2016MNRAS.457.2480C, 2017MNRAS.470.1750I, 2021A&A...650A.152I, 2024arXiv240810206L}. However, it is still not clear what mechanism(s) drive the instabilities. A number of processes have been proposed, such as disk turbulence \citep[e.g.][]{2012MNRAS.427L..21R,
2017AJ....153..120B}, perturbations from the oblateness of the rapidly-rotating host star \citep{2018AJ....155..167S, 2020AJ....160..105S}, overstable librations from eccentricity damping \citep{2014AJ....147...32G, 2015ApJ...810..119D, 2022ApJ...925...38N}, perturbations from distant giant planets \citep{2021MNRAS.508..597P,
2023A&A...674A.178B,
2023ApJ...954..137S}, secondary resonances between a fraction of the synodic frequency and
the MMR libration frequencies \citep{2020MNRAS.494.4950P}, or atmospheric mass loss from the planets \citep{matsumoto2020breaking}. Fortunately, the mechanism of the instability does not sensitively influence the post-instability outcomes, so it suffices to just pick one of these mechanisms. Following \citetalias{2022AJ....163..201G}, we consider atmospheric mass loss and impose exponential decay of each planet's mass. We select a damping timescale $\tau_m$ such that each planet loses 10\% of its mass over the subsequent 1 Myr following the initial 125 kyr period of resonant chain formation. 

\begin{figure}
    \centering
    \includegraphics[scale=0.5]{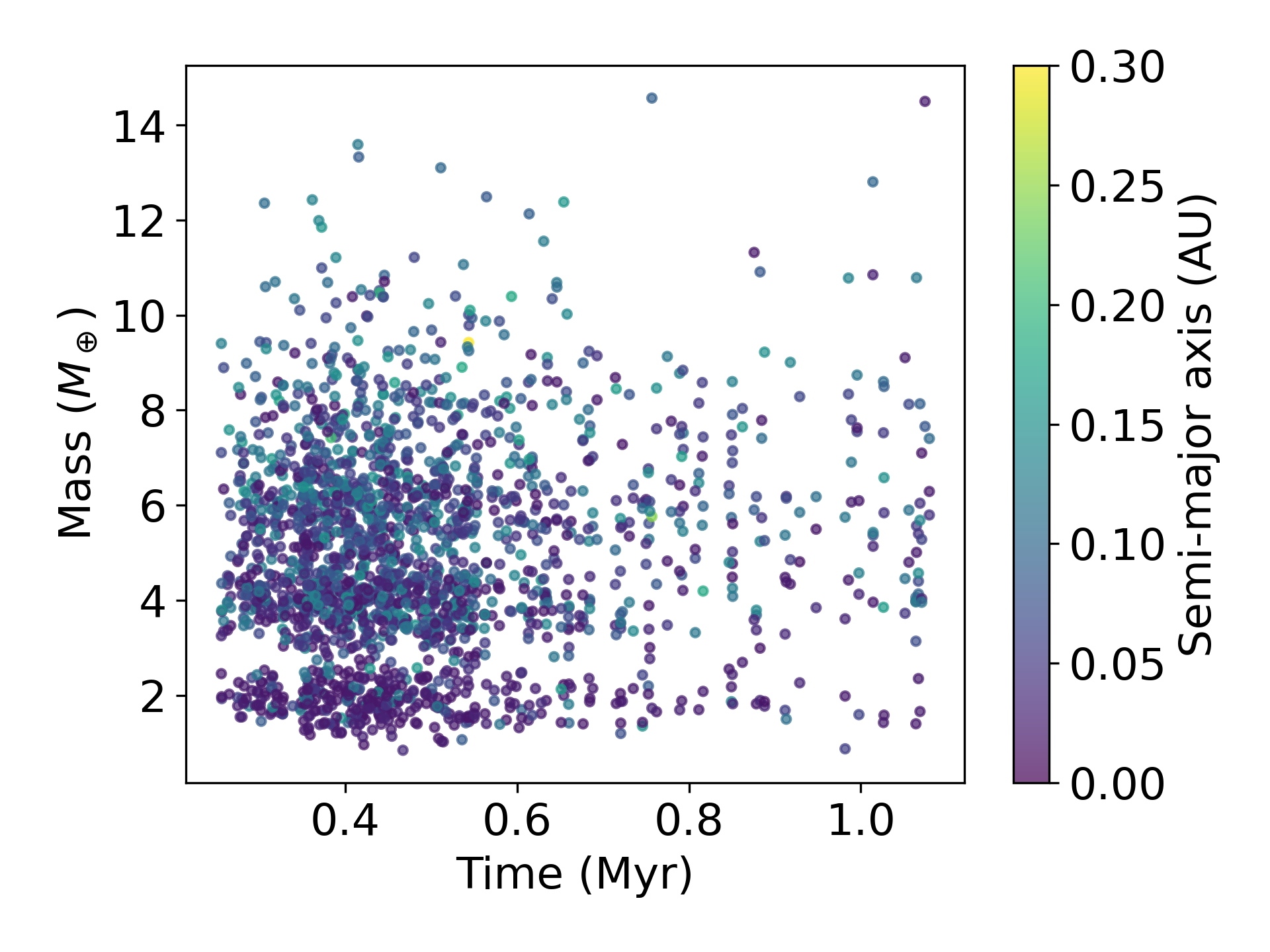}
    \caption{System properties following the last instability. We plot the masses of the final systems versus the time at which they undergo the final collision. The color indicates each planet's semi-major axis. All instabilities take place during the period of atmospheric mass loss (125 kyr $-$ 1.125 Myr.)}
    \label{fig:final_instabilites}
\end{figure}

Collisions are handled by adding the masses and conserving momentum of the collided planets. To confirm that the simulation timescale is sufficient for observing a majority of potential dynamical instabilities, we plot the instability events for our ``low-mass sample'' (to be described in Section \ref{sec: low-mass sample}) in \Cref{fig:final_instabilites} and notice that most occur within 0.5 Myr of the end of the migration period. Moreover, we use \texttt{SPOCK}, a deep learning tool which has been trained to predict dynamical instability probabilities and timescales \citep{tamayo2020predicting}, to evaluate the stability of each system after the simulation is complete. We observe that the vast majority of systems have predicted instability timescales exceeding 100 Myr, confirming that our simulation timescale is sufficient for observing most instabilities the systems will undergo. A summary of the full simulation sequence is provided in \Cref{table:pipeline}.

\section{Validation by Comparison to \citetalias{2022AJ....163..201G}}
\label{sec: high-mass sample}
%{\color{green} validate the simulation by running the high-mass simulation presented in Goldberg and Batygin and comparing the results}

\begin{table}
    \centering
    \begin{tabularx}{0.8\columnwidth}{c|c}
         Simulation phase & Time period\\
         \hline
         Disk migration & 0 kyr $-$ 100 kyr\\
         Disk dissipation & 100 kyr $-$ 110 kyr\\
         Monitoring period & 110 kyr $-$ 125 kyr\\
         Atmospheric mass loss & 125 kyr $-$ 1.125 Myr
    \end{tabularx}
    \caption{Summary of the simulation procedure.}
    \label{table:pipeline}
\end{table}

\begin{table*}
\begin{tabularx}{\textwidth}{ccccccccc}
Run & $M_\star \ [M_\odot]$ & initial $\overline{m} \ [M_\oplus]$ & final $\overline{m} \ [M_\oplus]$ & initial MMR & initial $\mathcal{D}$ & final $\mathcal{D}$ & final $r$ & final $\overline{\Delta}$ \\ \hline
\citetalias{2022AJ....163..201G} Run 1 & 1 & 8.0 & 16.6 & 4:3 & 0.00 & 0.31 & 0.34 & 10.2  \\ %\hline
High-mass sample & 1 & 8.0 & 17.1 & 4:3 & 0.00 & 0.33 & 0.42 & 15.5 \\ %\hline
Low-mass sample & 0.11 & 2.0 & 4.92 & 4:3 & 0.20 & 0.35 & 0.42 & 12.8 
\end{tabularx}
\caption{Comparison of the initial conditions and final statistics of simulations presented in \citetalias{2022AJ....163..201G} and in this work.}
\label{table: simulation summary}
\end{table*}

\begin{figure}
    \centering
    \includegraphics[scale=0.5]{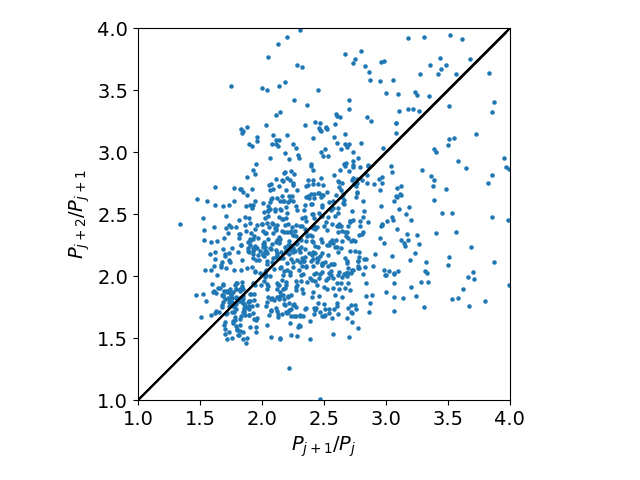}
    \caption{Period ratios between successive planets in each system in the high-mass sample.} %{\color{red} (This is definitely the high-mass sample, right?) I moved the figure because it was originally after Figure 5.}}
    \label{fig:period_ratios}
\end{figure}

We consider two different setups of the simulation procedure. The first is a validation simulation to compare to the results of \cite{2022AJ....163..201G} and confirm that we see similar statistical outcomes (this section). The second is the main focus of this study, that is collisional growth of planets orbiting very low-mass stars (Section \ref{sec: low-mass sample}).

Here we replicate one of the simulation sets presented in \citetalias{2022AJ....163..201G} and demonstrate agreement in intra-system uniformity statistics of the post-instability systems. We choose Run 1 of \citetalias{2022AJ....163..201G}, setting up systems with average planet mass $\overline{m} = 8 \ M_\oplus$ and dispersion $\mathcal{D} = 0$, orbiting a star of mass $1 \ M_\odot$. Furthermore, we choose $a_s = 0.1$ AU to reproduce the timescales used in \citetalias{2022AJ....163..201G}. The innermost planet is initialized at a semi-major axis of 0.1 AU and migrates inwards to around 0.09 AU due to interactions with the other planets. Each system is evolved according to the procedure outlined in Table \ref{table:pipeline}, although we perform an additional 5 Myr of evolution after atmospheric mass loss to better compare to \citetalias{2022AJ....163..201G}. We collect data for $100$ simulated systems. We refer to this as the ``high-mass sample''. 

We summarize the results of the simulations in Table \ref{table: simulation summary}, using a few metrics explored in \citetalias{2022AJ....163..201G}. For the collection of post-instability systems, we report a mass dispersion of $\mathcal{D} = 0.33$, showing agreement with Table 2 of \citetalias{2022AJ....163..201G}.

To further compare with previous literature \citep{pu2015spacing, weiss2018california} and demonstrate the long-term stability of the remnant systems, we compute the mutual Hill radius, given by
\begin{equation}
R_{Hj} = \left(\frac{m_{j+1}+m_j}{3M_{\star}}\right)^{1/3} \frac{a_{j+1} + a_j}{2},
\end{equation}
where $a_j$ and $m_j$ represent the semi-major axis and mass of the $j^\text{th}$ planet, and $M_\star$ is the stellar mass. The Hill spacing is defined as 
\begin{equation}
\Delta_j = \frac{a_{j+1}-a_j}{R_{Hj}}.
\end{equation}
Our simulations yield an average Hill spacing of $\overline{\Delta} = 15.5$, which slightly exceeds the value of $\overline{\Delta} = 10.2$ reported in \citetalias{2022AJ....163..201G}, but is still indicative of long-term stability of the post-instability systems \citep{pu2015spacing}.

Lastly, we test a metric of intra-system uniformity in our simulated systems. For each system, we compute the period ratio of successive planets as $P_{j+1} / P_{j}$. For each successive triplet of planets, we plot $P_{j+2} / P_{j+1}$ versus $P_{j+1} / P_{j}$ and carry out a least-squares regression (\Cref{fig:period_ratios}). We report a Pearson correlation coefficient of $r=0.42$, in agreement with studies of both real and synthetic planetary systems \citep{weiss2018california, 2022AJ....163..201G}.

\section{Collisional Growth of Planets around Low-Mass Stars}
\label{sec: low-mass sample}
%{\color{green} analyze the results of the low-mass simulation and compare/contrast with Golberg/Batygin}

\begin{figure*}
    \centering  \includegraphics[width=\textwidth]{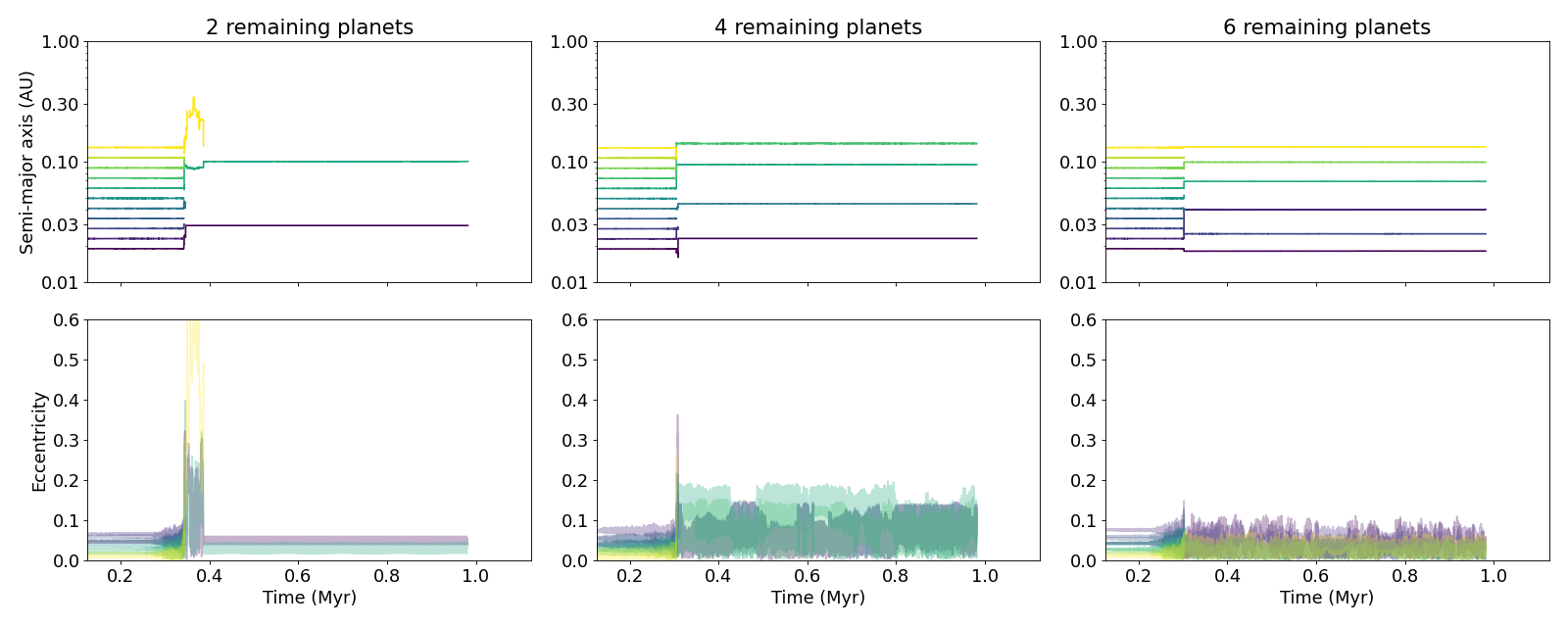}
    \caption{Semi-major axis and eccentricity evolution during the post-migration period (125 kyr - 1.125 Myr) for three systems in the low-mass sample. Each system represents a unique collisional outcome. }
    \label{fig:instability_period_1}
\end{figure*}

The results of the previous section showed that our simulations accurately reproduce the dynamics of instabilities in resonant chains orbiting a solar mass star. Here we extend our investigation to the regime of small planets orbiting very low-mass stars, with the aim of roughly reproducing the properties of LHS 3154 \citep{2023Sci...382.1031S}. We consider systems with a host star of mass $0.11 \ M_\odot$ (same as LHS 3154) and average planet mass $\overline{m} = 2 \ M_\oplus$.\footnote{Further discussion of this choice and the built-in assumptions can be found in Section \ref{sec: discussion}.} The innermost planet is initialized with semi-major axis of 0.04 AU, and we set the scaling parameter $a_s = 0.002$ AU. These parameters are chosen such that the post-migration system qualitatively agrees with observed systems of resonant super-Earths/sub-Neptunes orbiting very low-mass M dwarfs, such as TRAPPIST-1. We also use an initial mass dispersion of $\mathcal{D} = 0.2$, which represents a realistic level of intra-system uniformity \citep{2017ApJ...849L..33M}. Each planet's mass is drawn randomly from a Gaussian distribution with standard deviation determined by $\mathcal{D} = 0.2$.

We first analyze our suite of low-mass simulations in terms of the same statistical metrics discussed in the context of the high-mass sample. We compute the mass dispersion for each system, as well as the average mass dispersion $\mathcal{D}$. While each system's specific mass dispersion depends on which planets collide and thus varies with the initial conditions, we find an average final mass dispersion of $\mathcal{D}=0.35$, which agrees with our prior simulations of the high-mass sample and with the values reported in \citetalias{2022AJ....163..201G}. This supports \citetalias{2022AJ....163..201G}'s finding that the initial value of $\mathcal{D}$ does not significantly affect the outcomes of the planetary systems. We report a Pearson correlation coefficient of 0.42, indicating a similar level of intra-system uniformity as the simulations of the high-mass sample discussed earlier. We report an average Hill spacing of $\overline{\Delta} = 12.8$, consistent with long-term stability of the post-instability systems. %Furthermore, we report demonstrates that the average predicted instability timescale according to SPOCK \cite{tamayo2020predicting} exceeds {\color{red} todo Gyr} for most systems, indicating our choice of a {\color{red} todo Gyr} instability monitoring period is sufficient for this sample.

\begin{figure}
     \centering
     \includegraphics[width=\columnwidth]{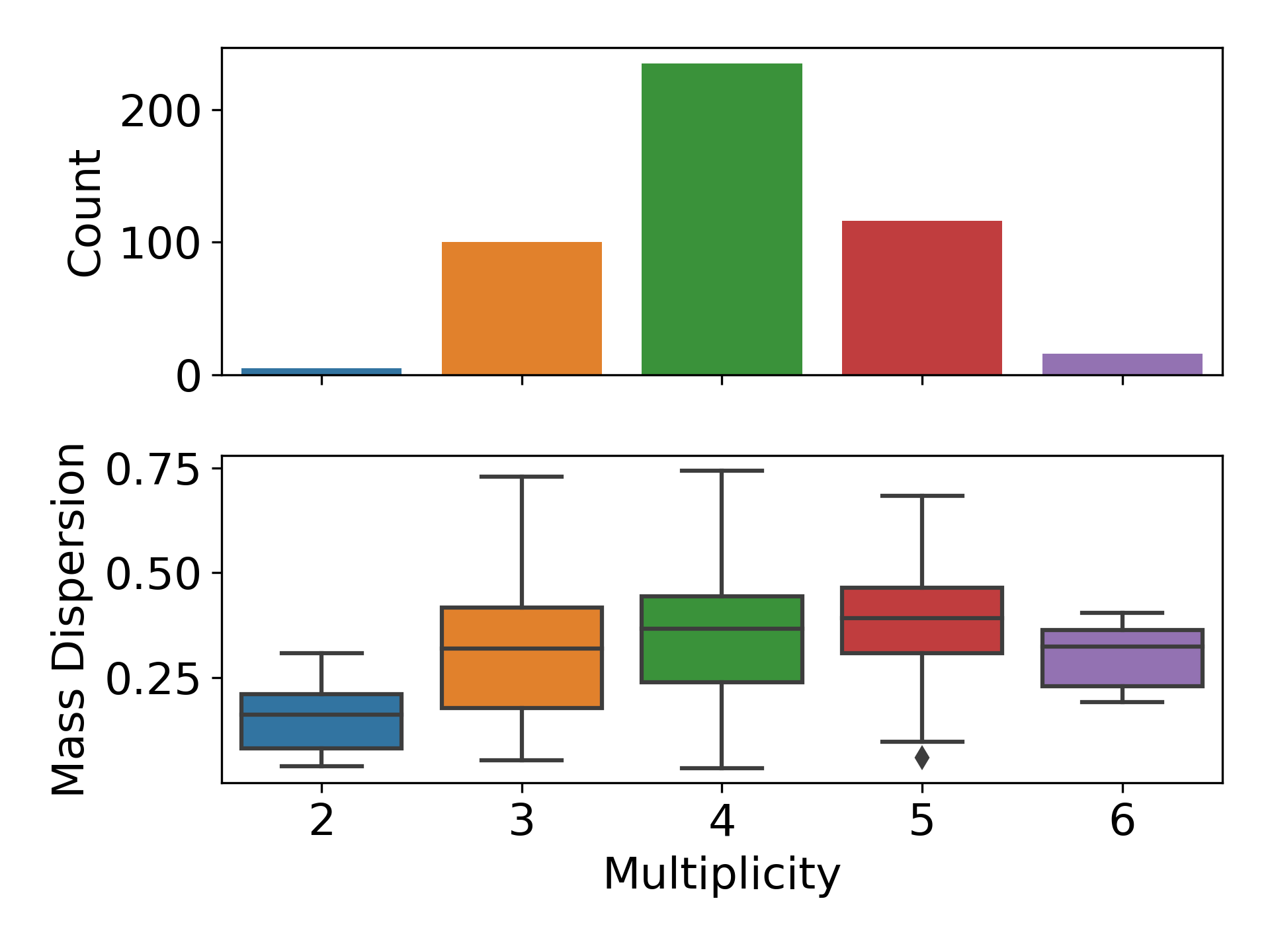}
     \caption{Top: Final multiplicities for post-instability systems in the low-mass sample. Bottom: Corresponding box plot distribution of mass dispersion for systems of each multiplicity. The middle three horizontal bars for each multiplicity indicate the first quartile, median, and third quartile of the data. One outlier, whose distance from the median exceeds $1.5$ times the interquartile range, is denoted with a diamond.} 
     \label{fig:mass_disp}
\end{figure}

\subsection{Characterization of instabilities}
%{\color{green} discuss plots which show collisions over time, etc. (collision energies?) might also be a good place to discuss the linear ``lower bound" for the period of high-mass planets}

We now proceed to analyze the instability outcomes, focusing on the low-mass sample. In this section we present a statistical summary of the collisional evolution across all simulations, before narrowing in on the close-in Neptune-mass planets in the next section. Systems typically undergo one or two instability events, each of which contain two to four collisions. Figure \ref{fig:instability_period_1} shows the orbital evolution of representative examples with two, four, and six remaining planets. Systems usually experience instabilities during the period of atmospheric mass dissipation; in contrast, for the high-mass sample, most instabilities occur $1-2$ Myr after atmospheric dissipation concludes. We hypothesize that the faster instability timescale in the low-mass sample is a result of a slightly larger planet-to-star mass ratio, which increases the relative strength of planet-planet interactions. 

All systems experienced at least one instability. Notably, all but one of the resulting systems had multiplicity $\leq 6$. In the subsequent analysis, we neglect the single system that only underwent a single collision during the integration. The final multiplicities and average mass dispersion of the low-mass sample are reported in \Cref{fig:mass_disp}. In agreement with the conclusions reported in \cite{10.1093/mnras/sty3463}, most remaining planets are the result of two to three collisions. A small proportion of systems yielded more massive planets which resulted from five or more collisions. Such cases will be discussed further in the next section. The mass dispersion does not vary significantly with the final multiplicity.

\begin{figure}
    \includegraphics[width=\columnwidth]{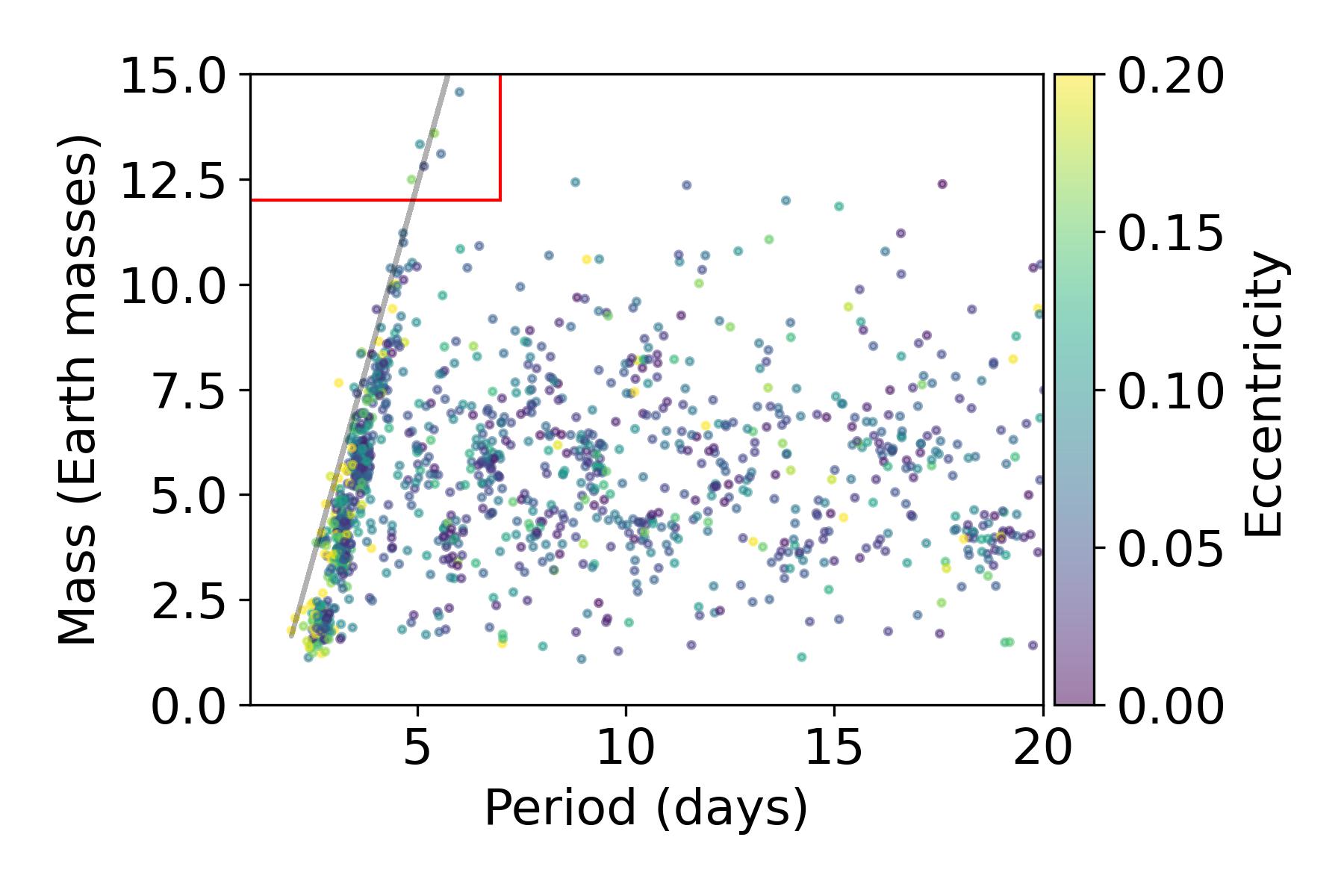}
    \caption{Mass and period of planets at the end of the simulations. Planets within the red box correspond to close-in Neptunes.}
    \label{fig:mass-period}
\end{figure}

\Cref{fig:mass-period} depicts the final orbital periods and masses of planets arising in the low-mass sample. Most planets have masses between $\sim 2-10 \ M_{\oplus}$, but there are a small number of planets more massive than $10 \ M_{\oplus}$. Notably, there appears to be a linear lower bound in period-mass space described by (for $P \gtrsim 1.5$ days)
\begin{equation}
\frac{m(P)}{M_{\oplus}} \approx -5.438 + 3.556 \left(\frac{P}{\mathrm{days}}\right).
\end{equation}
This can be explained qualitatively by analysis of the time-series data (\Cref{fig:instability_period_1}), where the remnant planet of a collision tends to have a higher average semi-major axis. Thus, we hypothesize that planets with shorter initial orbital periods tend to migrate outwards due to collisions with other planets, causing a deficiency in high-mass closely orbiting planets.

\subsection{Analysis of post-collision remnant planets}

\begin{figure}
    \centering
    \includegraphics[width=\columnwidth]{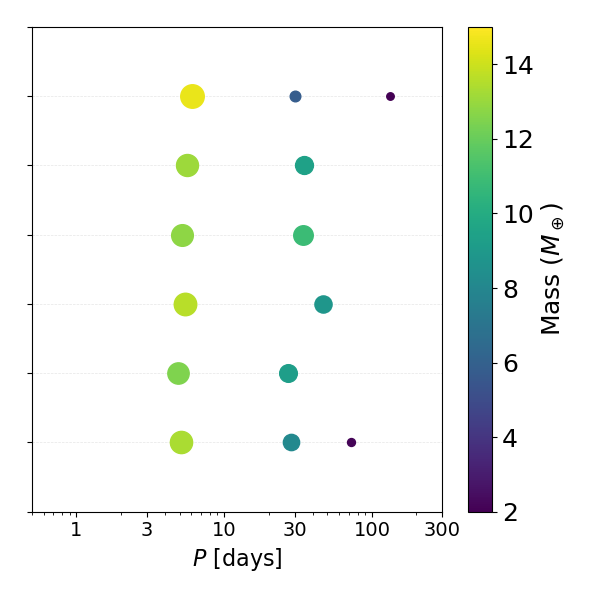}
    \caption{Architectures of post-instability systems for the six systems containing close-in Neptunes. The area is proportional to the square of the planet's inferred radius, based on the mass-radius relation of \cite{otegi2020revisited}. Each point is colored according to the planet's mass.}
    \label{fig:system_architectures}
\end{figure}

%{\color{green} discuss planet multiplicities and provide statistics on the probability of forming close-in Neptunes around ultracool star}

%(*In this section we can narrow in on the question of whether the simulations produce remnant exoplanets that resemble the close-in Neptune around an ultracool star*)

With a clear picture of the overall collision statistics, we can finally now address the question of whether the simulations produce close-in Neptune-like planets around their low-mass stellar hosts. Motivated by LHS 3154 b, we classify a planet as a close-in Neptune if it has $P < 7$ days and $M_p \in [12 \ M_\oplus, \ 20 \ M_\oplus]$. 

We find six close-in Neptunes across all 1,926 remnant planets. They are found in $1.2\%$ of systems. We note that the formation of a $> 10 \ M_\oplus$ planet requires no less than five collisions in our model, and thus close-in Neptunes tend to form in systems with low final multiplicity. The architectures of the six systems with close-in Neptunes are shown in Figure \ref{fig:system_architectures}. Four of these systems have two remaining planets, and two have three remaining planets. The close-in Neptune is always the innermost planet. Interestingly, the planets that are adjacent to them (i.e. the ``second'' planets) 
tend to have orbital periods around $\sim 30$ days. The period ratios between planets 1 and 2 are thus fairly large, $P_2/P_1 \sim 6-10$. This implies that the second planets would be more challenging to detect \citep[e.g.][]{2022AJ....164...72M}. However, searching for evidence of these outer companion planets offers a possible test of this model. \\

\section{Discussion}
\label{sec: discussion}

\subsection{Model assumptions}
\label{sec: assumptions}

Starting with $\sim 2 \ M_{\oplus}$ planets orbiting in a resonant chain around a very low-mass star like TRAPPIST-1, we showed that instabilities can drive multiple collisions and create a remnant close-in Neptune-like planet with a mass ratio $M_p/M_{\star} \sim 3 \times 10^{-4}$. However, our simulations made a number of assumptions. Here we discuss the validity of these assumptions and how they could be updated in future work. 

First, we assumed that the collisions resulted in perfect mergers. This is not true in general, since collisions can cause atmospheric mass loss \citep[e.g.][]{leinhardt2011collisions, 2019MNRAS.485.4454B, 2020MNRAS.496.1166D}. To assess our use of perfect mergers, we compute the ratio of the relative impact velocity and escape velocity of colliding planets. We report the mean, 99th percentile, and standard deviation of the velocity ratios as 0.46, 0.77, and 0.082. We note that for a vast majority of collision events, the impact velocities are less than the escape velocity of the resulting planets, indicating that our consideration of collisions as perfect mergers was reasonable. However, accounting for mass loss would be a possible future update to the model. 

Another assumption was the initial, pre-instability system architecture: 11 planets with $\sim2 \ M_{\oplus}$ masses in a chain of 4:3 MMRs. We already discussed in Section \ref{sec: methods} that a variety of first-order and second-order resonances are expected in reality. We also expect that not all systems would form long resonant chains. Although we are not modeling this diversity, our results are not strongly sensitive to this assumption.

Moreover, our set-up raises the question of whether $\sim22 \ M_{\oplus}$ in total initial solid mass is reasonable. To address this, we consider the \cite{burn2021new} study of planet formation around low-mass stars and explore their observationally-informed disk parameters. For stars of mass $\sim0.1 \ M_{\odot}$, they assumed approximate distributions of gas masses $\log_{10}(M_{\mathrm{gas}}/M_{\odot}) \sim \mathcal{N}(-2.5, 0.3)$ and dust-to-gas ratios $\log_{10}(f_{D/G}) \sim \mathcal{N}(-1.85, 0.22)$. The resulting mass in solids is approximately $\log_{10}(M_{\mathrm{solids}}/M_{\oplus}) \sim \mathcal{N}(1.17, 0.37)$, which yields a $1\sigma$ range of $\sim[6.3 \ M_{\oplus}, \ 35 \ M_{\oplus}]$ and a $3\sigma$ range of $\sim[1.1 \ M_{\oplus}, \ 191 \ M_{\oplus}]$. Our initial conditions imply a solid mass budget that is consistent with this distribution but on its upper end.

\subsection{Long-term tidal evolution}

Our simulations only modeled the first few Myr of evolution. The planets will continue to evolve over the subsequent billions of years, and it is these more evolved systems that are the analogs to most observed systems. Over these timescales, tides could substantially modify the orbits. We use some order-of-magnitude calculations to determine the impacts of tides. 

First, we note that the six remaining close-in Neptunes produced in the simulations in Section \ref{sec: low-mass sample} are left with moderate eccentricities in the range of $0.04-0.16$. We compute the timescale of semi-major axis decay due to eccentricity tides via equation 3 of \cite{levrard2009falling}. We neglect the tidal forces acting within the host star and make the simplifying assumption that the spin frequency is equal to the mean motion $n$ to obtain
\begin{equation}
\tau_t = \frac{a}{\dot{a}} \approx \frac{Q}{114 \pi} \frac{M_p}{M_{\star}}\left(\frac{ a}{R_p}\right)^5 \frac{P}{e^2},
\end{equation}
where $Q$ is the planet's tidal quality factor. The planet's radius is inferred from a mass-radius relation \citep{otegi2020revisited} and $Q$ is varied from $10^2$ to $10^4$.

For $Q=10^2$, we report tidal migration timescales of order $\sim 10$ Myr to $\sim 100$ Myr across the six remaining close-in Neptunes. For $Q$ ranging from $10^2$ to $10^4$, the tidal migration timescales range from $\sim 10$ Myr to $\sim 10$ Gyr, depending on the chosen quality factor. Due to the relatively low eccentricities, we expect that angular momentum-conserving tidal decay would only lead to a small decrease in orbital period of $3-20$\%. However, the presence of other planets in the systems may lead to angular momentum exchange and sustained eccentricity excitation of the close-in Neptunes, potentially amplifying the long-term effects of tidal dissipation \citep{pu2019low}. Thus, tides may induce substantial inward migration of the close-in Neptunes during their host star's lifetime, driving their orbital periods closer to that of LHS 3154 b. Long-term secular simulations could be performed to evaluate the extent of this migration.

\subsection{Comparison to other formation pathways}

\cite{2023Sci...382.1031S} explored core accretion and gravitational instability as two possible formation pathways for the formation of close-in Neptunes like LHS 3154 b. They found that extreme conditions within the core accretion scenario are potentially viable. Core accretion is the process by which protoplanets gather dust and gas mass from their protoplanetary disks \citep[e.g.][]{alibert2017formation, burn2021new}, but the outcome is dependent on the solid mass budget. \cite{miguel2020diverse} reported that the limited mass of protoplanetary disks for very low-mass stars would limit the resulting planets to $\lesssim 5 \ M_\oplus$. This makes LHS 3154 b impossible to explain without some modification. \cite{2023Sci...382.1031S} showed that planets as large as LHS 3154 b could be produced with an order-of-magnitude increase in the disk mass or dust-to-gas-ratio over that which is typically assumed for a $\sim0.1 \ M_{\odot}$ star. 

%The other potential formation pathway suggested by \cite{2023Sci...382.1031S} is gravitational instability. In this theory, the disks present near the beginning of the M dwarf's lifetime is relatively large ($\gtrsim 0.1\, M_\star$) and allows for the formation of sub-Jupiters \cite{mercer2020planet}. For $a < 100 \text{ AU}$, these disks can be unstable which leads to the formation of widely-orbiting ($\gtrsim 200$ days) gas giants ($\gtrsim 150 \, M_\oplus$.) However, for a star of mass $0.1 \, M_\odot$, numerical simulations have demonstrated that the minimum planet mass formed through this mechanism is ~60 earth masses, which greatly exceeds that of LHS 3154b \cite{morales2019giant}. To form a planet similar to LHS 3154 in this model, we would require an even higher protoplanetary disk mass than in the core accretion pathway.

%The struggles of both of these theories on sub-Neptune formation are of a similar vein -- given an ultracool star similar to LHS 3154, the mass present in the star's protoplanetary disk is insufficient to generate a close-in Neptune. 

%In particular, Stefannson et al. suggest that protoplanetary disks of low mass stars may grow to centimeter sizes, making them undetectable by milimeter observations. Alternatively, protoplanets may form much earlier than expected ($\ll 1 \text{ Myr})$, when the inferred masses of protoplanetary disks are much larger, allowing for core accretion to form a sub-Neptune.

Similar to \cite{2023Sci...382.1031S}, our hypothesized model requires a disk that is more massive than typical, but the required values fall within a physically reasonable range (Section \ref{sec: assumptions}). Our model leverages the well-studied phenomenon of dynamical instabilities arising in resonant chains. As we've demonstrated, these instabilities can culminate in the formation of close-in Neptunes similar to LHS 3154 b. In future work, it would be valuable to investigate ways of discriminating between these two models. One test would be to search for companion planets. \cite{2023Sci...382.1031S} predict no other planets nearby, and while we also predict no massive companion planets in very close proximity, we predict smaller planets in more distant orbits (Figure \ref{fig:system_architectures}).

%Resonant chains of super-Earths, such as TRAPPIST-1 and HD-110067, are believed to be a common outcome of protoplanetary disk migration, with the early onset of dynamical instabilities a primary explanation of their deficiency among observed systems \citep{lammers2024six, luger2017seven}.

\section{Conclusion}
\label{sec: conclusion}

Planets like LHS 3154 b pose a puzzle for traditional planet formation theories that predict only terrestrial planets around such tiny stars. Here we explored whether Neptune-mass planets around very low-mass stars may be a rare byproduct of the ``breaking the chains'' model of compact multi-planet system formation, in which small planets initially form in tightly-spaced resonant chains before subsequently experiencing dynamical instabilities and mergers. In part, this hypothesis was motivated by the existence of high-multiplicity systems of compact Earth/super-Earth-mass planets around low-mass stars (e.g. TRAPPIST-1), which could be consolidated into fairly massive planets in the event of a significant number of mergers.

We performed a suite of simulations of planets with initial average masses equal to $2 \ M_{\oplus}$ in tightly-spaced resonant chains around stars of mass $0.11 \ M_{\odot}$ (same as LHS 3154). The planets were subject to disk migration and mass loss. After a period of instabilities, the population broadly matches the properties of observed compact multi-planet systems, but alongside the high-multiplicity systems are a small number of systems that experienced $\geq 5$ collisions and left only two or three remaining planets. Close-in massive planets with $M_p \in [12 \ M_{\oplus}, \ 20 \ M_{\oplus}]$ and $P < 7$ days are produced in $\sim1$\% of systems. These massive planets are also left on slightly eccentric orbits ($e \sim 0.04 - 0.16$) that will cause their orbits to shrink by $\sim 3-20$\% (or perhaps more if there are strong secular planet-planet interactions). Moreover, the close-in Neptunes in our simulations are all accompanied by smaller planets at $\sim30$ days (Figure \ref{fig:system_architectures}).

While our theory is not the only possible explanation for LHS 3154 b, its benefit is mainly that the massive planets are a ``tail end of the Gaussian'' outcome of a preexisting model that already describes the rest of the planet population very well. Future observations can help test this theory. First, further refinement of occurrence rates of close-in Neptune-mass planets around $\lesssim 0.3 \ M_{\odot}$ stars would clarify just how rare they are and whether the theoretically predicted rate ($\sim1\%$) is consistent. Second, constraints on the properties of nearby companion planets would also be helpful. An anomalous set of mergers would leave a single massive planet with only smaller planets (if any) nearby (Figure \ref{fig:system_architectures}). With M dwarfs currently being ideal targets to search for temperate rocky planets, we expect future observations to gradually reveal a more complete picture of their system architectures. \\

\section{Acknowledgements}
We thank Gu\dh mundur Stef\'ansson for helpful comments on a draft of the manuscript. We also thank the anonymous reviewer for their thorough and insightful comments. We thank Greg Laughlin and Christopher Spalding for helpful conversations. This material is based upon work supported by the National Science Foundation under Grant No. 2306391. We also acknowledge support from the MIT Undergraduate Research Opportunities Program. We gratefully acknowledge access to computational resources through the MIT Engaging cluster at the Massachusetts Green High Performance Computing Center (MGHPCC) facility and the MIT SuperCloud and Lincoln Laboratory Supercomputing Center \citep{reuther2018interactive}. 

\bibliographystyle{aasjournal}
\bibliography{main}

\begin{thebibliography}{}
\expandafter\ifx\csname natexlab\endcsname\relax\def\natexlab#1{#1}\fi
\providecommand{\url}[1]{\href{#1}{#1}}
\providecommand{\dodoi}[1]{doi:~\href{http://doi.org/#1}{\nolinkurl{#1}}}
\providecommand{\doeprint}[1]{\href{http://ascl.net/#1}{\nolinkurl{http://ascl.net/#1}}}
\providecommand{\doarXiv}[1]{\href{https://arxiv.org/abs/#1}{\nolinkurl{https://arxiv.org/abs/#1}}}

\bibitem[{{Agol} {et~al.}(2021){Agol}, {Dorn}, {Grimm}, {Turbet}, {Ducrot},
  {Delrez}, {Gillon}, {Demory}, {Burdanov}, {Barkaoui}, {Benkhaldoun},
  {Bolmont}, {Burgasser}, {Carey}, {de Wit}, {Fabrycky}, {Foreman-Mackey},
  {Haldemann}, {Hernandez}, {Ingalls}, {Jehin}, {Langford}, {Leconte},
  {Lederer}, {Luger}, {Malhotra}, {Meadows}, {Morris}, {Pozuelos}, {Queloz},
  {Raymond}, {Selsis}, {Sestovic}, {Triaud}, \& {Van
  Grootel}}]{2021PSJ.....2....1A}
{Agol}, E., {Dorn}, C., {Grimm}, S.~L., {et~al.} 2021, \psj, 2, 1,
  \dodoi{10.3847/PSJ/abd022}

\bibitem[{Alibert \& Benz(2017)}]{alibert2017formation}
Alibert, Y., \& Benz, W. 2017, Astronomy \& Astrophysics, 598, L5

\bibitem[{{Almenara} {et~al.}(2024){Almenara}, {Bonfils}, {Bryant},
  {Jord{\'a}n}, {H{\'e}brard}, {Martioli}, {Correia}, {Astudillo-Defru},
  {Cadieux}, {Arnold}, {Artigau}, {Bakos}, {Barros}, {Bayliss}, {Bouchy},
  {Bou{\'e}}, {Brahm}, {Carmona}, {Charbonneau}, {Ciardi}, {Cloutier},
  {Cointepas}, {Cook}, {Cowan}, {Delfosse}, {Dias do Nascimento}, {Donati},
  {Doyon}, {Forveille}, {Fouqu{\'e}}, {Gaidos}, {Gilbert}, {Gomes da Silva},
  {Hartman}, {Hesse}, {Hobson}, {Jenkins}, {Kiefer}, {Kostov}, {Laskar},
  {Lendl}, {L'Heureux}, {Martins}, {Menou}, {Moutou}, {Murgas}, {Polanski},
  {Rapetti}, {Sedaghati}, \& {Shang}}]{2024A&A...683A.166A}
{Almenara}, J.~M., {Bonfils}, X., {Bryant}, E.~M., {et~al.} 2024, \aap, 683,
  A166, \dodoi{10.1051/0004-6361/202346999}

\bibitem[{{Batygin} \& {Adams}(2017)}]{2017AJ....153..120B}
{Batygin}, K., \& {Adams}, F.~C. 2017, \aj, 153, 120,
  \dodoi{10.3847/1538-3881/153/3/120}

\bibitem[{{Biersteker} \& {Schlichting}(2019)}]{2019MNRAS.485.4454B}
{Biersteker}, J.~B., \& {Schlichting}, H.~E. 2019, \mnras, 485, 4454,
  \dodoi{10.1093/mnras/stz738}

\bibitem[{{Bitsch} \& {Izidoro}(2023)}]{2023A&A...674A.178B}
{Bitsch}, B., \& {Izidoro}, A. 2023, \aap, 674, A178,
  \dodoi{10.1051/0004-6361/202245040}

\bibitem[{{Bonfils} {et~al.}(2013){Bonfils}, {Delfosse}, {Udry}, {Forveille},
  {Mayor}, {Perrier}, {Bouchy}, {Gillon}, {Lovis}, {Pepe}, {Queloz}, {Santos},
  {S{\'e}gransan}, \& {Bertaux}}]{2013A&A...549A.109B}
{Bonfils}, X., {Delfosse}, X., {Udry}, S., {et~al.} 2013, \aap, 549, A109,
  \dodoi{10.1051/0004-6361/201014704}

\bibitem[{{Bryant} {et~al.}(2023){Bryant}, {Bayliss}, \& {Van
  Eylen}}]{2023MNRAS.521.3663B}
{Bryant}, E.~M., {Bayliss}, D., \& {Van Eylen}, V. 2023, \mnras, 521, 3663,
  \dodoi{10.1093/mnras/stad626}

\bibitem[{Burn {et~al.}(2021)Burn, Schlecker, Mordasini, Emsenhuber, Alibert,
  Henning, Klahr, \& Benz}]{burn2021new}
Burn, R., Schlecker, M., Mordasini, C., {et~al.} 2021, Astronomy \&
  Astrophysics, 656, A72

\bibitem[{{Ca{\~n}as} {et~al.}(2023){Ca{\~n}as}, {Kanodia}, {Libby-Roberts},
  {Lin}, {Schutte}, {Powers}, {Jones}, {Monson}, {Wang}, {Stef{\'a}nsson},
  {Cochran}, {Robertson}, {Mahadevan}, {Kowalski}, {Wisniewski}, {Parker},
  {Larsen}, {Chapman}, {Kobulnicky}, {Gupta}, {Everett}, {Penprase}, {Zeimann},
  {Beard}, {Bender}, {Col{\'o}n}, {Diddams}, {Fredrick}, {Halverson}, {Ninan},
  {Ramsey}, {Roy}, \& {Schwab}}]{2023AJ....166...30C}
{Ca{\~n}as}, C.~I., {Kanodia}, S., {Libby-Roberts}, J., {et~al.} 2023, \aj,
  166, 30, \dodoi{10.3847/1538-3881/acdac7}

\bibitem[{{Carrera} {et~al.}(2019){Carrera}, {Ford}, \&
  {Izidoro}}]{2019MNRAS.486.3874C}
{Carrera}, D., {Ford}, E.~B., \& {Izidoro}, A. 2019, \mnras, 486, 3874,
  \dodoi{10.1093/mnras/stz974}

\bibitem[{{Coleman} \& {Nelson}(2016)}]{2016MNRAS.457.2480C}
{Coleman}, G. A.~L., \& {Nelson}, R.~P. 2016, \mnras, 457, 2480,
  \dodoi{10.1093/mnras/stw149}

\bibitem[{{Cresswell} \& {Nelson}(2006)}]{2006A&A...450..833C}
{Cresswell}, P., \& {Nelson}, R.~P. 2006, \aap, 450, 833,
  \dodoi{10.1051/0004-6361:20054551}

\bibitem[{{Dai} {et~al.}(2024){Dai}, {Goldberg}, {Batygin}, {van Saders},
  {Chiang}, {Choksi}, {Li}, {Petigura}, {Gilbert}, {Millholland}, {Dai},
  {Bouma}, {Weiss}, \& {Winn}}]{2024arXiv240606885D}
{Dai}, F., {Goldberg}, M., {Batygin}, K., {et~al.} 2024, arXiv e-prints,
  arXiv:2406.06885, \dodoi{10.48550/arXiv.2406.06885}

\bibitem[{{Deck} \& {Batygin}(2015)}]{2015ApJ...810..119D}
{Deck}, K.~M., \& {Batygin}, K. 2015, \apj, 810, 119,
  \dodoi{10.1088/0004-637X/810/2/119}

\bibitem[{{Denman} {et~al.}(2020){Denman}, {Leinhardt}, {Carter}, \&
  {Mordasini}}]{2020MNRAS.496.1166D}
{Denman}, T.~R., {Leinhardt}, Z.~M., {Carter}, P.~J., \& {Mordasini}, C. 2020,
  \mnras, 496, 1166, \dodoi{10.1093/mnras/staa1623}

\bibitem[{{Emsenhuber} {et~al.}(2021{\natexlab{a}}){Emsenhuber}, {Mordasini},
  {Burn}, {Alibert}, {Benz}, \& {Asphaug}}]{2021A&A...656A..69E}
{Emsenhuber}, A., {Mordasini}, C., {Burn}, R., {et~al.} 2021{\natexlab{a}},
  \aap, 656, A69, \dodoi{10.1051/0004-6361/202038553}

\bibitem[{{Emsenhuber} {et~al.}(2021{\natexlab{b}}){Emsenhuber}, {Mordasini},
  {Burn}, {Alibert}, {Benz}, \& {Asphaug}}]{2021A&A...656A..70E}
---. 2021{\natexlab{b}}, \aap, 656, A70, \dodoi{10.1051/0004-6361/202038863}

\bibitem[{{Endl} {et~al.}(2006){Endl}, {Cochran}, {K{\"u}rster}, {Paulson},
  {Wittenmyer}, {MacQueen}, \& {Tull}}]{2006ApJ...649..436E}
{Endl}, M., {Cochran}, W.~D., {K{\"u}rster}, M., {et~al.} 2006, \apj, 649, 436,
  \dodoi{10.1086/506465}

\bibitem[{{Esteves} {et~al.}(2022){Esteves}, {Izidoro}, {Bitsch}, {Jacobson},
  {Raymond}, {Deienno}, \& {Winter}}]{2022MNRAS.509.2856E}
{Esteves}, L., {Izidoro}, A., {Bitsch}, B., {et~al.} 2022, \mnras, 509, 2856,
  \dodoi{10.1093/mnras/stab3203}

\bibitem[{{Gan} {et~al.}(2022){Gan}, {Lin}, {Wang}, {Mao}, {Fouqu{\'e}}, {Fan},
  {Bedell}, {Stassun}, {Giacalone}, {Fukui}, {Murgas}, {Ciardi}, {Howell},
  {Collins}, {Shporer}, {Arnold}, {Barclay}, {Charbonneau}, {Christiansen},
  {Crossfield}, {Dressing}, {Elliott}, {Esparza-Borges}, {Evans}, {Gnilka},
  {Gonzales}, {Howard}, {Isogai}, {Kawauchi}, {Kurita}, {Liu}, {Livingston},
  {Matson}, {Narita}, {Palle}, {Parviainen}, {Rackham}, {Rodriguez}, {Rose},
  {Rudat}, {Schlieder}, {Scott}, {Vezie}, {Ricker}, {Vanderspek}, {Latham},
  {Seager}, {Winn}, \& {Jenkins}}]{2022MNRAS.511...83G}
{Gan}, T., {Lin}, Z., {Wang}, S.~X., {et~al.} 2022, \mnras, 511, 83,
  \dodoi{10.1093/mnras/stab3708}

\bibitem[{{Gan} {et~al.}(2023){Gan}, {Wang}, {Wang}, {Mao}, {Huang}, {Collins},
  {Stassun}, {Shporer}, {Zhu}, {Ricker}, {Vanderspek}, {Latham}, {Seager},
  {Winn}, {Jenkins}, {Barkaoui}, {Belinski}, {Ciardi}, {Evans}, {Girardin},
  {Maslennikova}, {Mazeh}, {Panahi}, {Pozuelos}, {Radford}, {Schwarz},
  {Twicken}, {W{\"u}nsche}, \& {Zucker}}]{2023AJ....165...17G}
{Gan}, T., {Wang}, S.~X., {Wang}, S., {et~al.} 2023, \aj, 165, 17,
  \dodoi{10.3847/1538-3881/ac9b12}

\bibitem[{{Gillon} {et~al.}(2017){Gillon}, {Triaud}, {Demory}, {Jehin}, {Agol},
  {Deck}, {Lederer}, {de Wit}, {Burdanov}, {Ingalls}, {Bolmont}, {Leconte},
  {Raymond}, {Selsis}, {Turbet}, {Barkaoui}, {Burgasser}, {Burleigh}, {Carey},
  {Chaushev}, {Copperwheat}, {Delrez}, {Fernandes}, {Holdsworth}, {Kotze}, {Van
  Grootel}, {Almleaky}, {Benkhaldoun}, {Magain}, \&
  {Queloz}}]{2017Natur.542..456G}
{Gillon}, M., {Triaud}, A. H.~M.~J., {Demory}, B.-O., {et~al.} 2017, \nat, 542,
  456, \dodoi{10.1038/nature21360}

\bibitem[{{Goldberg} \& {Batygin}(2022)}]{2022AJ....163..201G}
{Goldberg}, M., \& {Batygin}, K. 2022, \aj, 163, 201,
  \dodoi{10.3847/1538-3881/ac5961}

\bibitem[{{Goldreich} \& {Schlichting}(2014)}]{2014AJ....147...32G}
{Goldreich}, P., \& {Schlichting}, H.~E. 2014, \aj, 147, 32,
  \dodoi{10.1088/0004-6256/147/2/32}

\bibitem[{Grimm {et~al.}(2018)Grimm, Demory, Gillon, Dorn, Agol, Burdanov,
  Delrez, Sestovic, Triaud, Turbet, {et~al.}}]{grimm2018nature}
Grimm, S.~L., Demory, B.-O., Gillon, M., {et~al.} 2018, Astronomy \&
  Astrophysics, 613, A68

\bibitem[{{He} {et~al.}(2019){He}, {Ford}, \&
  {Ragozzine}}]{2019MNRAS.490.4575H}
{He}, M.~Y., {Ford}, E.~B., \& {Ragozzine}, D. 2019, \mnras, 490, 4575,
  \dodoi{10.1093/mnras/stz2869}

\bibitem[{{Ida} \& {Lin}(2005)}]{2005ApJ...626.1045I}
{Ida}, S., \& {Lin}, D.~N.~C. 2005, \apj, 626, 1045, \dodoi{10.1086/429953}

\bibitem[{{Izidoro} {et~al.}(2021){Izidoro}, {Bitsch}, {Raymond}, {Johansen},
  {Morbidelli}, {Lambrechts}, \& {Jacobson}}]{2021A&A...650A.152I}
{Izidoro}, A., {Bitsch}, B., {Raymond}, S.~N., {et~al.} 2021, \aap, 650, A152,
  \dodoi{10.1051/0004-6361/201935336}

\bibitem[{{Izidoro} {et~al.}(2017){Izidoro}, {Ogihara}, {Raymond},
  {Morbidelli}, {Pierens}, {Bitsch}, {Cossou}, \&
  {Hersant}}]{2017MNRAS.470.1750I}
{Izidoro}, A., {Ogihara}, M., {Raymond}, S.~N., {et~al.} 2017, \mnras, 470,
  1750, \dodoi{10.1093/mnras/stx1232}

\bibitem[{{Johnson} {et~al.}(2007){Johnson}, {Butler}, {Marcy}, {Fischer},
  {Vogt}, {Wright}, \& {Peek}}]{2007ApJ...670..833J}
{Johnson}, J.~A., {Butler}, R.~P., {Marcy}, G.~W., {et~al.} 2007, \apj, 670,
  833, \dodoi{10.1086/521720}

\bibitem[{{Kanodia} {et~al.}(2023){Kanodia}, {Mahadevan}, {Libby-Roberts},
  {Stefansson}, {Ca{\~n}as}, {Piette}, {Boss}, {Teske}, {Chambers}, {Zeimann},
  {Monson}, {Robertson}, {Ninan}, {Lin}, {Bender}, {Cochran}, {Diddams},
  {Gupta}, {Halverson}, {Hawley}, {Kobulnicky}, {Metcalf}, {Parker}, {Powers},
  {Ramsey}, {Roy}, {Schwab}, {Swaby}, {Terrien}, \&
  {Wisniewski}}]{2023AJ....165..120K}
{Kanodia}, S., {Mahadevan}, S., {Libby-Roberts}, J., {et~al.} 2023, \aj, 165,
  120, \dodoi{10.3847/1538-3881/acabce}

\bibitem[{{Kanodia} {et~al.}(2024){Kanodia}, {Ca{\~n}as}, {Mahadevan}, {Ford},
  {Helled}, {Anderson}, {Boss}, {Cochran}, {Delamer}, {Han}, {Libby-Roberts},
  {Lin}, {M{\"u}ller}, {Robertson}, {Stef{\'a}nsson}, \&
  {Teske}}]{2024AJ....167..161K}
{Kanodia}, S., {Ca{\~n}as}, C.~I., {Mahadevan}, S., {et~al.} 2024, \aj, 167,
  161, \dodoi{10.3847/1538-3881/ad27cb}

\bibitem[{{Laughlin} {et~al.}(2004){Laughlin}, {Bodenheimer}, \&
  {Adams}}]{2004ApJ...612L..73L}
{Laughlin}, G., {Bodenheimer}, P., \& {Adams}, F.~C. 2004, \apjl, 612, L73,
  \dodoi{10.1086/424384}

\bibitem[{Leinhardt \& Stewart(2011)}]{leinhardt2011collisions}
Leinhardt, Z.~M., \& Stewart, S.~T. 2011, The Astrophysical Journal, 745, 79

\bibitem[{{Leleu} {et~al.}(2024){Leleu}, {Delisle}, {Burn}, {Izidoro}, {Udry},
  {Dumusque}, {Lovis}, {Millholland}, {Parc}, {Bouchy}, {Bourrier}, {Alibert},
  {Faria}, {Mordasini}, \& {S{\'e}gransan}}]{2024A&A...687L...1L}
{Leleu}, A., {Delisle}, J.-B., {Burn}, R., {et~al.} 2024, \aap, 687, L1,
  \dodoi{10.1051/0004-6361/202450587}

\bibitem[{Levrard {et~al.}(2009)Levrard, Winisdoerffer, \&
  Chabrier}]{levrard2009falling}
Levrard, B., Winisdoerffer, C., \& Chabrier, G. 2009, The Astrophysical
  Journal, 692, L9

\bibitem[{{Li} {et~al.}(2024){Li}, {Chiang}, {Choksi}, \&
  {Dai}}]{2024arXiv240810206L}
{Li}, R., {Chiang}, E., {Choksi}, N., \& {Dai}, F. 2024, arXiv e-prints,
  arXiv:2408.10206.
\newblock \doarXiv{2408.10206}

\bibitem[{Luger {et~al.}(2017)Luger, Sestovic, Kruse, Grimm, Demory, Agol,
  Bolmont, Fabrycky, Fernandes, Van~Grootel, {et~al.}}]{luger2017seven}
Luger, R., Sestovic, M., Kruse, E., {et~al.} 2017, Nature Astronomy, 1, 0129

\bibitem[{Matsumoto \& Ogihara(2020)}]{matsumoto2020breaking}
Matsumoto, Y., \& Ogihara, M. 2020, The Astrophysical Journal, 893, 43

\bibitem[{{Ment} \& {Charbonneau}(2023)}]{2023AJ....165..265M}
{Ment}, K., \& {Charbonneau}, D. 2023, \aj, 165, 265,
  \dodoi{10.3847/1538-3881/acd175}

\bibitem[{Miguel {et~al.}(2020)Miguel, Cridland, Ormel, Fortney, \&
  Ida}]{miguel2020diverse}
Miguel, Y., Cridland, A., Ormel, C., Fortney, J., \& Ida, S. 2020, Monthly
  Notices of the Royal Astronomical Society, 491, 1998

\bibitem[{{Miguel} {et~al.}(2020){Miguel}, {Cridland}, {Ormel}, {Fortney}, \&
  {Ida}}]{2020MNRAS.491.1998M}
{Miguel}, Y., {Cridland}, A., {Ormel}, C.~W., {Fortney}, J.~J., \& {Ida}, S.
  2020, \mnras, 491, 1998, \dodoi{10.1093/mnras/stz3007}

\bibitem[{{Millholland} {et~al.}(2017){Millholland}, {Wang}, \&
  {Laughlin}}]{2017ApJ...849L..33M}
{Millholland}, S., {Wang}, S., \& {Laughlin}, G. 2017, \apjl, 849, L33,
  \dodoi{10.3847/2041-8213/aa9714}

\bibitem[{{Millholland} {et~al.}(2022){Millholland}, {He}, \&
  {Zink}}]{2022AJ....164...72M}
{Millholland}, S.~C., {He}, M.~Y., \& {Zink}, J.~K. 2022, \aj, 164, 72,
  \dodoi{10.3847/1538-3881/ac7c67}

\bibitem[{{Nesvorn{\'y}} {et~al.}(2022){Nesvorn{\'y}}, {Chrenko}, \&
  {Flock}}]{2022ApJ...925...38N}
{Nesvorn{\'y}}, D., {Chrenko}, O., \& {Flock}, M. 2022, \apj, 925, 38,
  \dodoi{10.3847/1538-4357/ac36cd}

\bibitem[{{Ogihara} \& {Ida}(2009)}]{2009ApJ...699..824O}
{Ogihara}, M., \& {Ida}, S. 2009, \apj, 699, 824,
  \dodoi{10.1088/0004-637X/699/1/824}

\bibitem[{Otegi {et~al.}(2020)Otegi, Bouchy, \& Helled}]{otegi2020revisited}
Otegi, J., Bouchy, F., \& Helled, R. 2020, Astronomy \& Astrophysics, 634, A43

\bibitem[{{Pass} {et~al.}(2023){Pass}, {Winters}, {Charbonneau}, {Irwin},
  {Latham}, {Berlind}, {Calkins}, {Esquerdo}, \& {Mink}}]{2023AJ....166...11P}
{Pass}, E.~K., {Winters}, J.~G., {Charbonneau}, D., {et~al.} 2023, \aj, 166,
  11, \dodoi{10.3847/1538-3881/acd349}

\bibitem[{{Pichierri} \& {Morbidelli}(2020)}]{2020MNRAS.494.4950P}
{Pichierri}, G., \& {Morbidelli}, A. 2020, \mnras, 494, 4950,
  \dodoi{10.1093/mnras/staa1102}

\bibitem[{Pu \& Lai(2019)}]{pu2019low}
Pu, B., \& Lai, D. 2019, Monthly Notices of the Royal Astronomical Society,
  488, 3568

\bibitem[{{Pu} \& {Lai}(2021)}]{2021MNRAS.508..597P}
{Pu}, B., \& {Lai}, D. 2021, \mnras, 508, 597, \dodoi{10.1093/mnras/stab2504}

\bibitem[{Pu \& Wu(2015)}]{pu2015spacing}
Pu, B., \& Wu, Y. 2015, The Astrophysical Journal, 807, 44

\bibitem[{{Rein}(2012)}]{2012MNRAS.427L..21R}
{Rein}, H. 2012, \mnras, 427, L21, \dodoi{10.1111/j.1745-3933.2012.01337.x}

\bibitem[{{Rein} \& {Liu}(2012)}]{rebound}
{Rein}, H., \& {Liu}, S.~F. 2012, \aap, 537, A128,
  \dodoi{10.1051/0004-6361/201118085}

\bibitem[{{Rein} \& {Tamayo}(2015)}]{reboundwhfast}
{Rein}, H., \& {Tamayo}, D. 2015, \mnras, 452, 376,
  \dodoi{10.1093/mnras/stv1257}

\bibitem[{Reuther {et~al.}(2018)Reuther, Kepner, Byun, Samsi, Arcand, Bestor,
  Bergeron, Gadepally, Houle, Hubbell, Jones, Klein, Milechin, Mullen, Prout,
  Rosa, Yee, \& Michaleas}]{reuther2018interactive}
Reuther, A., Kepner, J., Byun, C., {et~al.} 2018, in 2018 IEEE High Performance
  extreme Computing Conference (HPEC), IEEE, 1--6

\bibitem[{{Sabotta} {et~al.}(2021){Sabotta}, {Schlecker}, {Chaturvedi},
  {Guenther}, {Mu{\~n}oz Rodr{\'\i}guez}, {Mu{\~n}oz S{\'a}nchez}, {Caballero},
  {Shan}, {Reffert}, {Ribas}, {Reiners}, {Hatzes}, {Amado}, {Klahr}, {Morales},
  {Quirrenbach}, {Henning}, {Dreizler}, {Pall{\'e}}, {Perger}, {Azzaro},
  {Jeffers}, {Kaminski}, {K{\"u}rster}, {Lafarga}, {Montes}, {Passegger}, \&
  {Zechmeister}}]{2021A&A...653A.114S}
{Sabotta}, S., {Schlecker}, M., {Chaturvedi}, P., {et~al.} 2021, \aap, 653,
  A114, \dodoi{10.1051/0004-6361/202140968}

\bibitem[{{Sobski} \& {Millholland}(2023)}]{2023ApJ...954..137S}
{Sobski}, N., \& {Millholland}, S.~C. 2023, \apj, 954, 137,
  \dodoi{10.3847/1538-4357/ace966}

\bibitem[{{Spalding} {et~al.}(2018){Spalding}, {Marx}, \&
  {Batygin}}]{2018AJ....155..167S}
{Spalding}, C., {Marx}, N.~W., \& {Batygin}, K. 2018, \aj, 155, 167,
  \dodoi{10.3847/1538-3881/aab43a}

\bibitem[{{Spalding} \& {Millholland}(2020)}]{2020AJ....160..105S}
{Spalding}, C., \& {Millholland}, S.~C. 2020, \aj, 160, 105,
  \dodoi{10.3847/1538-3881/aba629}

\bibitem[{{Stef{\'a}nsson} {et~al.}(2023){Stef{\'a}nsson}, {Mahadevan},
  {Miguel}, {Robertson}, {Delamer}, {Kanodia}, {Ca{\~n}as}, {Winn}, {Ninan},
  {Terrien}, {Holcomb}, {Ford}, {Zawadzki}, {Bowler}, {Bender}, {Cochran},
  {Diddams}, {Endl}, {Fredrick}, {Halverson}, {Hearty}, {Hill}, {Lin},
  {Metcalf}, {Monson}, {Ramsey}, {Roy}, {Schwab}, {Wright}, \&
  {Zeimann}}]{2023Sci...382.1031S}
{Stef{\'a}nsson}, G., {Mahadevan}, S., {Miguel}, Y., {et~al.} 2023, Science,
  382, 1031, \dodoi{10.1126/science.abo0233}

\bibitem[{{Tamayo} {et~al.}(2020){Tamayo}, {Rein}, {Shi}, \&
  {Hernandez}}]{reboundx}
{Tamayo}, D., {Rein}, H., {Shi}, P., \& {Hernandez}, D.~M. 2020, \mnras, 491,
  2885, \dodoi{10.1093/mnras/stz2870}

\bibitem[{Tamayo {et~al.}(2020)Tamayo, Cranmer, Hadden, Rein, Battaglia,
  Obertas, Armitage, Ho, Spergel, Gilbertson, {et~al.}}]{tamayo2020predicting}
Tamayo, D., Cranmer, M., Hadden, S., {et~al.} 2020, Proceedings of the National
  Academy of Sciences, 117, 18194

\bibitem[{{Terquem} \& {Papaloizou}(2007)}]{2007ApJ...654.1110T}
{Terquem}, C., \& {Papaloizou}, J. C.~B. 2007, \apj, 654, 1110,
  \dodoi{10.1086/509497}

\bibitem[{{Weiss} {et~al.}(2023){Weiss}, {Millholland}, {Petigura}, {Adams},
  {Batygin}, {Block}, \& {Mordasini}}]{2023ASPC..534..863W}
{Weiss}, L.~M., {Millholland}, S.~C., {Petigura}, E.~A., {et~al.} 2023, in
  Astronomical Society of the Pacific Conference Series, Vol. 534, Protostars
  and Planets VII, ed. S.~{Inutsuka}, Y.~{Aikawa}, T.~{Muto}, K.~{Tomida}, \&
  M.~{Tamura}, 863

\bibitem[{Weiss {et~al.}(2018)Weiss, Marcy, Petigura, Fulton, Howard, Winn,
  Isaacson, Morton, Hirsch, Sinukoff, {et~al.}}]{weiss2018california}
Weiss, L.~M., Marcy, G.~W., Petigura, E.~A., {et~al.} 2018, The Astronomical
  Journal, 155, 48

\bibitem[{Wisdom \& Holman(1991)}]{wisdom1991symplectic}
Wisdom, J., \& Holman, M. 1991, Astronomical Journal (ISSN 0004-6256), vol.
  102, Oct. 1991, p. 1528-1538., 102, 1528

\bibitem[{Zink {et~al.}(2018)Zink, Christiansen, \&
  Hansen}]{10.1093/mnras/sty3463}
Zink, J.~K., Christiansen, J.~L., \& Hansen, B. M.~S. 2018, Monthly Notices of
  the Royal Astronomical Society, 483, 4479, \dodoi{10.1093/mnras/sty3463}

\end{thebibliography}

\end{document}